\documentclass{pasj01}
\Received{$\langle$reception date$\rangle$}
\Accepted{$\langle$acception date$\rangle$}
\Published{$\langle$publication date$\rangle$}

\usepackage{amsthm}

\usepackage{color}
\newcommand{\ba}{\begin{eqnarray}}
\newcommand{\ea}{\end{eqnarray}}

\newcommand{\jcal}{J2216+3518 }
\newcommand{\kms}{km s$^{-1}$}

\begin{document}

\title{Annual parallax measurements of a semiregular variable star SV Pegasus with VERA}
\author{Hiroshi Sudou,\altaffilmark{1,*}  Toshihiro Omodaka, \altaffilmark{2} Kotone Murakami,\altaffilmark{2} Takahiro Nagayama, \altaffilmark{2} Akiharu Nakagawa, \altaffilmark{2}  Riku Urago, \altaffilmark{2} Takumi Nagayama, \altaffilmark{3} Ken Hirano,  \altaffilmark{3} Mareki Honma \altaffilmark{3} }
\altaffiltext{1}{Faculty of Engineering, Gifu University, 1-1 Yanagido, Gifu, Gifu 501-1193}
\altaffiltext{2}{Faculty of Science, Kagoshima University, 21-24 Korimoto, Kagoshima, Kagoshima 890-8580}
\altaffiltext{3}{Mizusawa VLBI Observatory, National Astronomical Observatory of Japan, 2-12 Hoshigaoka-cho, Mizusawa, Oshu, Iwate 023-0861}
\email{sudou@gifu-u.ac.jp}


\maketitle

\begin{abstract}

Many studies have shown that there are clear sequences in the period-luminosity relationship (PLR) for Mira variables and semiregular variables (SRVs) in the Large Magellanic Cloud (LMC). To investigate the PLR for SRVs in our galaxy, we examined the annual parallax measurement and conducted K'-band photometric monitoring of an SRV star SV Pegasus (SV Peg). 
We measured the position change of the associating H$_2$O maser spots by phase-referencing VLBI observations with VERA at 22 GHz, spanning approximately 3 yr, and detected an annual parallax of $\pi = 3.00 \pm 0.06$ mas, corresponding to a distance of $D=333 \pm 7$ pc. This result is in good agreement with the Hipparcos parallax and improves the accuracy of the distance from 35 $\%$ to 2 $\%$. However, the GAIA DR2 catalog gave a parallax of $\pi=1.12\pm0.28$ mas for SV Peg. This indicates that the GAIA result might be blurred by the effect of the stellar size because the estimated stellar radius was $\sim 5$ mas, which is comparable to the parallax.
We obtained a K'-band mean magnitude of $m_{K'} = -0.48$ mag and a period of $P=177$ days from our photometric monitoring with a 1-m telescope. Using the trigonometric distance, we derived an absolute magnitude of $M_{K'}=-8.09 \pm 0.05$ mag. This result shows that the position of SV Peg in the PLR falls on the C' sequence found in the PLR in the LMC, which is similar to other SRVs in our galaxy. 


\end{abstract}

\section{Introduction}

It is important to clarify the period-luminosity relationship (PLR) of variable stars for the study of stellar evolution.
Because of the numerous efforts made to examine the relation between logarithm of pulsation period and K'-band magnitude using massive databases obtained in the Large Magellanic Cloud (LMC),
several clear sequences in the PLR have been found for Mira variables and semiregular variables (SRV) (Glass \& Evans 1981, Feast et al. 1989, Wood 2000, Glass \&  van Leeuwen 2007). 
These objects can be found in sequences designated as C and C', which are considered to be populated by the Mira variables pulsating in the fundamental mode and those pulsating in the first overtone mode, respectively (Ita et al. 2004a, b). 
If the PLR in the LMC were similar to that in our galaxy, it could be very useful to estimate the distance to variable stars in our galaxy once the pulsation period was observed.
However, it is not clear whether or not the LMC and our galaxy share the same PLR; it possibly depends on the difference in metallicity between them.
To establish the PLR in our galaxy, we need to measure both the distance and period to individual variable stars.  

In this study, we show the results of the monitoring observations of an SRV star SV Peg. 
SV Peg is classified as SRb of M7 spectral type, and its pulsation period was estimated to be 144.6 days by the General Catalog of Variable Stars (Samus' et al. 2017). 
This source has been known to exhibit a strong H$_2$O and SiO maser emission as observed in many Mira variables and SRVs (e.g., Kim et al. 2010, Shintani et al. 2008).  
Using phase-referencing VLBI observations of H$_2$O masers at 22 GHz, we measured the annual parallax of SV Peg. 
We also carried out Near Infrared (NIR) photometric monitoring to investigate the average apparent magnitude and the precise pulsation period. 
These results allow us to investigate the star's position in the PLR of our galaxy, proper motions of the star, and its stellar properties.

\section{Observations and Data Reduction}

\subsection{Single dish monitoring}

A time series of the H$_2$O maser spectra of SV Peg at 22 GHz was observed from June 2012 to April 2014, using the 20-m telescope at VERA IRIKI station in Japan. 
In total, we have obtained 12 maser spectra. 
The digital spectrometer installed at the telescope provides a velocity resolution of 0.05 km s$^{-1}$ per channel, and the spectral range covers 32 MHz (432 km s$^{-1}$). 
A standard on-off beam-switching technique was used for the observation of the source. 
The system noise temperatures were measured for each observation and were used to correct the antenna temperature 
for atmospheric attenuation and changes in antenna gain as a function of the elevation angle.
The typical amplitude error is estimated to be 5 -- 20 \% in our R-sky calibration system of the telescope (Shintani et al. 2008). 


\subsection{Near Infrared monitoring}

We carried out near-infrared observations of SV Peg from June 2012 to August 2016 using the Kagoshima University 1-m telescope. The near-infrared camera equipped with the telescope has a 512$\times$ 512 pixels HAWAII array, which provides the J-, H-, and K'-band images. An image scale of the
array is 0.636 pixel, yielding a field of view of 5.025 $\times$ 5.025 arcmin. The data reduction
and photometry for these data were carried out using the National Optical Astronomy Observatory Imaging Reduction and Facility (IRAF) software package. Standard procedures of data reduction were adopted. The average dark frame was subtracted, normalizing each of the dark-subtracted images by the flat-field frame. Subsequently, the sky frame was subtracted from the normalized image. The photometry was carried out with the IRAF/APPHOT package. Since SV Peg is too bright in the near-infrared and causes a saturation of the detector, we used two observation modes -- the image-defocus technique between 2012 and 2013, and a technique using the Local Attenuation Filter (LAF) between 2014 and 2016. LAF was originally developed for the IRSF 1.4-m telescope for the observation
of very bright NIR stars (Nagayama 2016) and was also installed on the Kagoshima
University 1-m telescope (Nagayama et al.  in prep).

\subsection{VLBI observations}

The H$_2$O maser emission at 22 GHz associated with SV Peg was observed using the VLBI Exploration of Radio Astrometry (VERA) between June 2012 and April 2015. VERA consists of four telescopes at Mizusawa, Iriki, Ogasawara, and Ishigaki-jima, and its minimum fringe spacing is 1 mas. A bright  extragalactic continuum source \jcal was observed simultaneously with a dual-beam system equipped with all the antennas of VERA. This source was used as both a position reference and a phase calibrator in our phase-referencing observations. The separation angle between SV Peg and \jcal is 2.17 $^\circ$. 
The detailed observations for each epoch are summarized in Table 1. 
The observation coordinates of SV Peg and \jcal were $(\alpha, \delta)_{\rm J2000}=$ (22h05m42.08385s, +35d20'54.5280") and (22h16m20.009903s, +35d18'14.18005") at the phase tracking center, respectively. 

The received signals were recorded using the Sony DIR 2000 system with a recording rate of 1 Gbps, with 16 video channels of 16-MHz bandwidth. 
Correlation processes have been carried out using the FX correlator at Mitaka. 
Each video band was divided into 512 frequency channels, corresponding to a velocity resolution of 0.4 \kms.

We used the NRAO AIPS package for fringe fitting and image synthesis. 
Successive velocity channels containing the strong maser spots were used to calibrate the fringe rates. 
The same channels were also used as fringe-phase references for a self-calibration procedure. 
Amplitude calibration was performed using the system noise temperatures during the observations. 
For phase-referencing, fringe fitting was performed using the AIPS task FRING on \jcal. 
The solutions were applied to the data of SV Peg to calibrate the visibility data. 
Phase and amplitude solutions obtained from self-calibration of \jcal were also applied to SV Peg.
The instrumental phase difference between the two beams was measured continuously during the observations by injecting artificial noise sources into both beams (Honma et al. 2008a). The tropospheric delay was calibrated based on GPS measurements at the zenith delay (Honma et al. 2008b). 
After the calibration, we made spectral-line image cubes using the AIPS task IMAGR around masers with arrays of 1024 $\times$ 1024 pixels, with a cell size of 0.05 mas.
The positions of the maser spots were estimated by using the AIPS task JMFIT, which is used for 2-dimensional Gaussian fits. 

During all epochs, \jcal showed a point-like structure (Figure 1), which is consistent with the previous result observed with VERA (Petrov et al. 2012). 
The time variation of the peak intensity of \jcal is summarized in Table 2. The change in the peak intensity of \jcal  is estimated to be 14 $\%$. These results indicate that \jcal is a very good source for phase-referencing. 

\section{Results}

\subsection{Time Variations of Masers}

Figure 2 shows the H$_2$O spectra of SV Peg with the single dish observations. 
We found three main components, peaking around $V_{\rm LSR}=3.6, 5.5$, and 6.8 \kms. 
These results are consistent with previous observations (e.g., Kim et al. 2010). 
The time variation of the integrated intensity of the spectrum is shown in Figure 3. The integrated intensity varies between 10 and 90 mJy \kms in our observational span. 
It was reported that the intensity variation during several years seems to be irregular and that the correlation between maser and optical data is almost negligible for SV Peg (Winnberg et al. 2008).

\subsection{Infrared Light curve}

Results of our photometric monitoring using the Kagoshima 1-m telescope are presented in Figure 4. Using K'-band data, we solved period $P$ by assuming a sinusoidal function of

\begin{eqnarray}
m = m_0 + A \sin \left( 2 \pi\frac{t+ \theta}{P} \right), 
\end{eqnarray}
where $m$ is the apparent magnitude, $m_0$ is the averaged magnitude, $A$ is amplitude, and $\theta$ is the voluntary phase. Details of the fitting procedure are shown in Nakagawa et al. (2016).

Since SRVs often show several periods, we present the fitting results with local minimum root-mean-square (RMS) in the limited time span. 
In Table 4 and Figure 3, we derived three possible periods: 177, 260, and 341 days. Please note that the longest value (341 days) is almost twice the shortest value (177 days).  Furthermore, we carried out the same fitting procedure using V-band data obtained by Hipparcos and derived a single period of 172 days (Figure 5). This value is similar to the period of 177 days obtained from the K'-band data. Therefore, we concluded that the pulsation period of SV Peg is 177 days. 
By applying this result, we derived the averaged K'-band magnitude of $-0.48\pm0.01$ mag. 

In the same way, we carried out fitting to J- and H-band data by fixing the period to 177 days (Figure 6). Although the data are slightly sparse, fitting seems to be reasonable. The fitting results are summarized in Table 3.

\subsection{Annual Parallax and the Distance to SV Peg}

From the phase-referencing map ranging 300 $\times$ 300 mas, we detected 53 water maser spots in all epochs (Table 4).  
In order to perform the astrometric analysis, we used the following selection criteria for the maser spots: (1) constant LSR velocity
 {\bf at the different epochs},
(2) similar proper motions, and (3) lifetimes longer than one year. 
Taking above criteria into account, among these 53 spots, we found 7 spots  
{\bf (Spot 1 -- 3, 6 -- 9 in Table 4)}  available for the annual parallax measurement (Table 5). 

Figure 7 shows the position changes of these 7 spots.
{\bf Using the position change of these spots, we carried out least-square fitting
 of a common annual parallax to these spots and individual proper motions with a constant velocity for each spot. }
We evaluated the position error of the maser spots by adding the formal error from the residual of Gaussian fitting by JMFIT and the systematic error to achieve the reduced $\chi^2$ of unity for the parallax fit, because the formal errors are often underestimated from the true position error  (e.g., Nagayama et al. 2015). 
The fitting result 
{\bf by using the 7 spots}
is shown in Figure 8. 
As a result, we estimated an annual parallax 
$\pi=3.00\pm0.06$ mas, corresponding to a distance $D=333\pm7$ pc.
This result is consistent with that obtained with Hipparcos ($\pi=4.28\pm1.49$ mas) and improves the accuracy of the distance from 35 $\%$ to 2 $\%$. 
Using the trigonometric distance, we derived an absolute magnitude $M_{K'}=-8.09 \pm 0.05$ mag. 

Since some maser spots could be detected in the same velocity channel in general, 
we would possibly select different combinations of the positions of the maser spots (Imai et al. 2007, Dzib et al. 2014).
Thus, we carried out a test to confirm how the annual parallax changes if we adopt the results by another two analysts independently, and found that the difference of the annual parallaxes among them is less than the fitting error of the parallax. 

\subsection{Proper Motion}

Assuming that the annual parallax is common among all maser spots, we obtained the proper motion for 29 spots detected for greater than 2 epochs. The averaged proper motion is estimated to be $11.59\pm0.54$ and $-8.63\pm0.44$ mas yr$^{-1}$ for the RA and Dec directions, respectively. These values are in good agreement with those previously obtained by Hipparcos, 12.27$\pm$1.17 and $-9.73\pm1.24$ mas yr$^{-1}$ (van Leeuwen 2007). Thus, we refer to the averaged proper motion we derived to be the systemic motion of SV Peg. 
This corresponds to $22.8\pm1.1$ \kms. 

\section{Discussion}

\subsection{Period-Magnitude Diagram}

Figure 9 shows the log $P$ - $M_{K}$ diagram in our galaxy including Mira variables and SRVs whose distances were determined with phase-referencing  VLBI methods (Nakagawa et al. 2016 and references therein). Note that Nakagawa et al. (2016) multiply the periods of SRVs by two to include them together with Mira variables in their fundamental tones, but we used the observed value directly. We found that SV Peg falls near the other three SRVs in the diagram, and these SRVs clearly fall on different regions from Mira variables in the diagram. 
We also show the defined sequences found in the LMC by Ita et al. (2004a, b), modified to be available in our galaxy (Kamezaki et al. 2014). 
SV Peg clearly falls on sequence C' (first overtone mode), as well as the other SRVs. 

\subsection{Stellar Properties}

To estimate the bolometric luminosity of SV Peg, we calculated the K'-band bolometric correction ($BC_K$). 
According to Bessel \& Wood (1984), the analytic fit to the bolometric correction for K-band is given by

\begin{eqnarray}
BC_K=0.60 + 2.65 (m_J - m_{K}) - 0.67 (m_J - m_{K})^2
\end{eqnarray}

where $0.6 \le (m_J - m_{K}) \le 2$. Using our J-band and K'-band mean magnitude with 1-m telescope, we obtained $BC_K=3.1$. This leads to a bolometric magnitude of 2.6 mag,
and a bolometric luminosity of $7.8 \times 10^3 L_{\odot}$. Applying the effective temperature of $2.9 \times 10^3$ K estimated from the ISO-SWS model fitting (Aringer et al. 2002), 
a stellar radius of 350 $R_{\odot}$ was obtained, corresponding to 4.8 mas for the distance of SV Peg.

\subsection{Distribution and Internal Motion of the Masers}

It is reasonable to suppose that the internal motion of the maser spots can be measured with respect to the systemic motion discussed in section 3.4. We show the internal motion and distribution of the maser spots in Figure 10 and Table 4. 
 The maser spots are distributed bimodally, at the northwest (NW) and southeast (SE). 
The NW and SE components tend to move opposite to each other, indicating possible bipolar outflows. Such behavior has often been reported in some SRVs (e.g., Ishitsuka et al. 2001). 

The deviation of the internal motion is estimated to be 2.2 and 1.5 mas yr$^{-1}$ (3.5 and 2.5 \kms) for the RA and Dec directions, respectively. These values are slightly larger than the dispersion of the radial velocity of 1.0 \kms. If the maser spots in SV Peg are in the bipolar outflow, expanding with constant velocity, its speed is calculated to be 4.4 \kms with an inclination angle of 80$^\circ$. Since CO(J=2-1) observations indicated the expanding velocity to be 7.5 \kms, the water maser shell is a likely place to be accelerated by the dust radiation's pressure in the circumstellar envelope.

\subsection{Comparison with GAIA parallax}

Recently, the GAIA DR2 catalog gave $\pi=1.12\pm0.28$ mas for the parallax of SV Peg (Gaia Collaboration et al.  2016, Gaia Collaboration et al. 2018), leading to a much larger distance than those in our result. 
Applying the GAIA parallax, we obtain $M_{K}\sim-10$ mag. This value seems to be much larger than the typical magnitude of SRVs in our galaxy (see Figure 9). 
We would like to note that this difference might be related to the size of the photosphere of SV Peg. 
 It is natural that the astrometry for resolved stars should show larger values of trigonometric parallaxes instead of smaller.
 However, our observation is not in agreement with this picture; i.e, the GAIA parallax is much smaller than the VLBI parallax we obtained, though the estimated size of $5$ mas is larger than the observed GAIA parallax of 1 mas. 
 We will discuss statistically the origin of the difference of the parallaxes between GAIA and VLBI  in the separated paper (Nakagawa et al. in prep.). 


\section{Conclusion}

We determined an annual parallax of SV Peg of $3.00\pm0.06$ mas, corresponding to a distance of $333\pm7$ pc by using phase-referencing VLBI observations. 
We obtained a K'-band mean magnitude of $m_{K'} = -0.48$ mag and a period of $P=177$ days from our photometric monitoring with the 1-m telescope. Using the determined distance, we derived the absolute magnitude of $M_{K'}=-8.09 \pm 0.05$ mag. This result suggests that the position of SV Peg in the PLR falls on the C' sequence found in the PLR in LMC. 
Based on the analysis of the internal motion of the maser spots, we found a possible bipolar outflow in the envelope in SV Peg. If we accept this picture, its projected speed is estimated to be $\sim$4 \kms.

\begin{ack}

We are grateful to the members of the VERA project for their kind collaboration and encouragement. 
This work has made use of data from the European Space Agency (ESA) mission
{\it Gaia} ({https://www.cosmos.esa.int/gaia}), processed by the {\it Gaia}
Data Processing and Analysis Consortium (DPAC,
{https://www.cosmos.esa.int/web/gaia/dpac/consortium}). Funding for the DPAC
has been provided by national institutions, in particular, the institutions
participating in the {\it Gaia} Multilateral Agreement.

\end{ack}

{}

\begin{figure}
 \begin{center}
  \includegraphics[width=8cm]{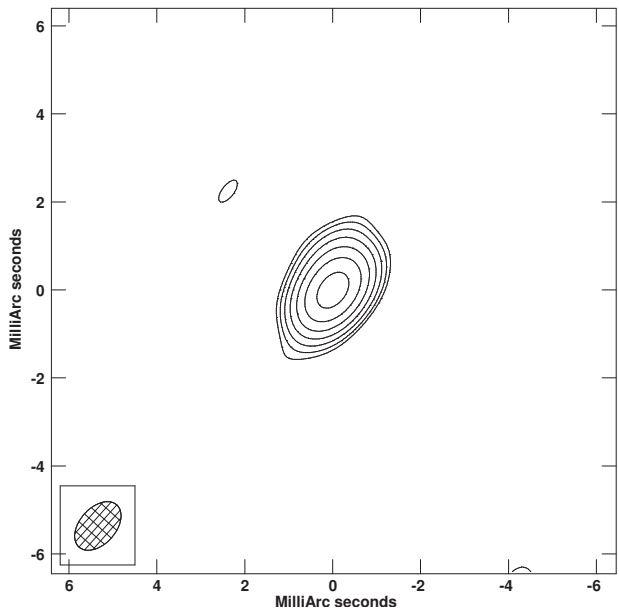}
 \end{center} \vspace{2cm}
 \caption{Image of the calibrator source 2216+3518. }\label{fig1}
\end{figure}
\begin{figure}
 \begin{center}
  \includegraphics[width=10cm]{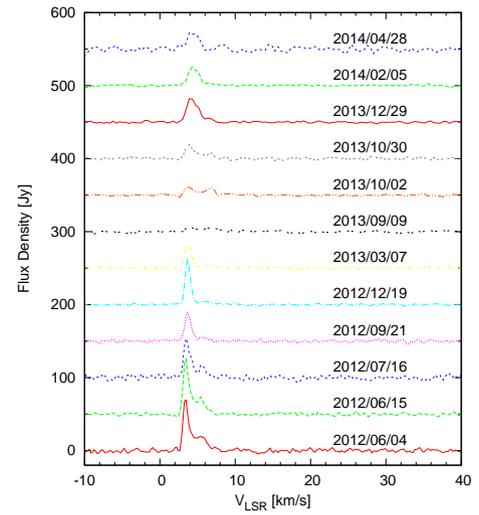}
 \end{center} \vspace{2cm}
 \caption{Spectra of H$_2$O maser emission in SV Peg.}\label{fig1}
\end{figure}
\begin{figure}
 \begin{center}
  \includegraphics[width=8cm]{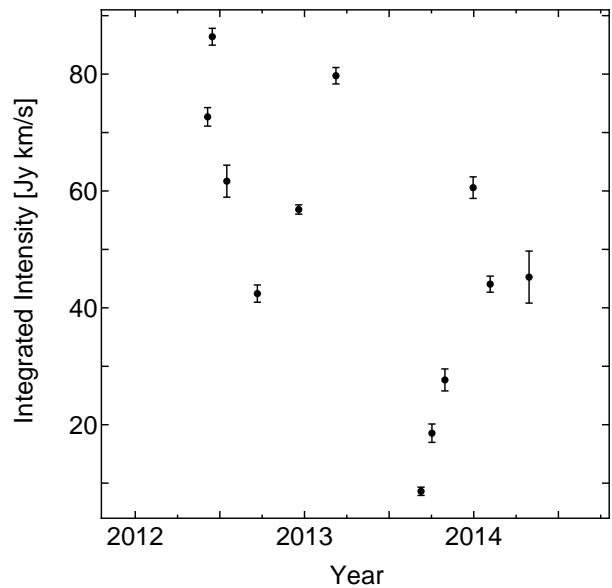}
 \end{center} \vspace{2cm}
 \caption{Time variation of the integrated intensity of H$_2$O maser emission in SV Peg. }\label{fig1}
\end{figure}
\begin{figure}
 \begin{center}
  \includegraphics[width=8cm]{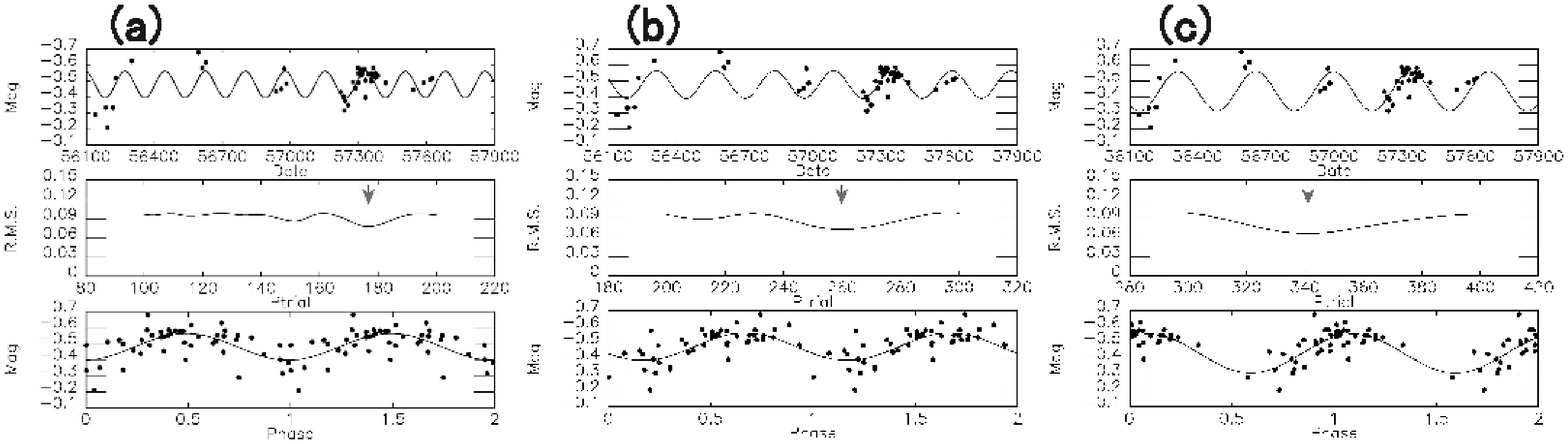}
 \end{center} \vspace{2cm}
 \caption{K'-band light curve and fitting results with sinusoidal function (top), the RMS of fitting residual (middle), and
phase-folded light curve with the fitted period (bottom).  (a) $P=177$ days, (b) $P=260$ days, and (c) $P=341$ days. }\label{fig1}
\end{figure}
\begin{figure}
 \begin{center}
  \includegraphics[width=8cm]{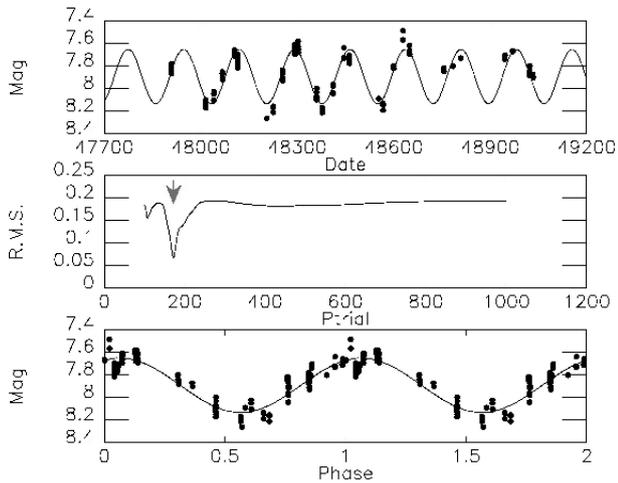}
 \end{center} \vspace{2cm}
 \caption{V-band light curve and fitting results with sinusoidal function (top), the RMS of fitting residual (middle), and
phase-folded light curve with the fitted period (bottom). }\label{fig1}
\end{figure}
\begin{figure}[r]
 \begin{center}
  \includegraphics[width=8cm]{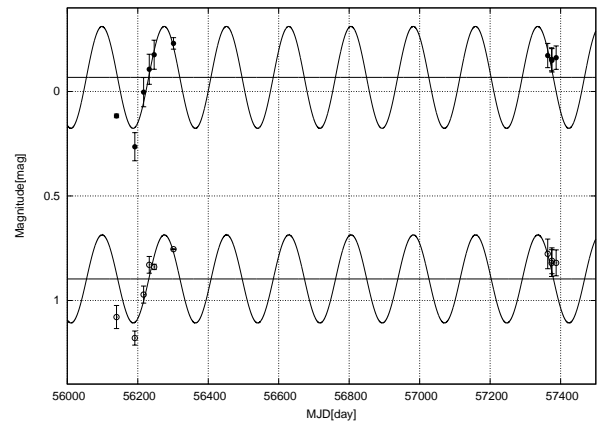}
 \end{center} \vspace{2cm}
 \caption{H- (filled circle) and J- (open circle) band light curve and fitting results with sinusoidal function by using a fixed period of 177 days.   }\label{fig1}
\end{figure}
\begin{figure}
 \begin{center}
  \includegraphics[width=5cm]{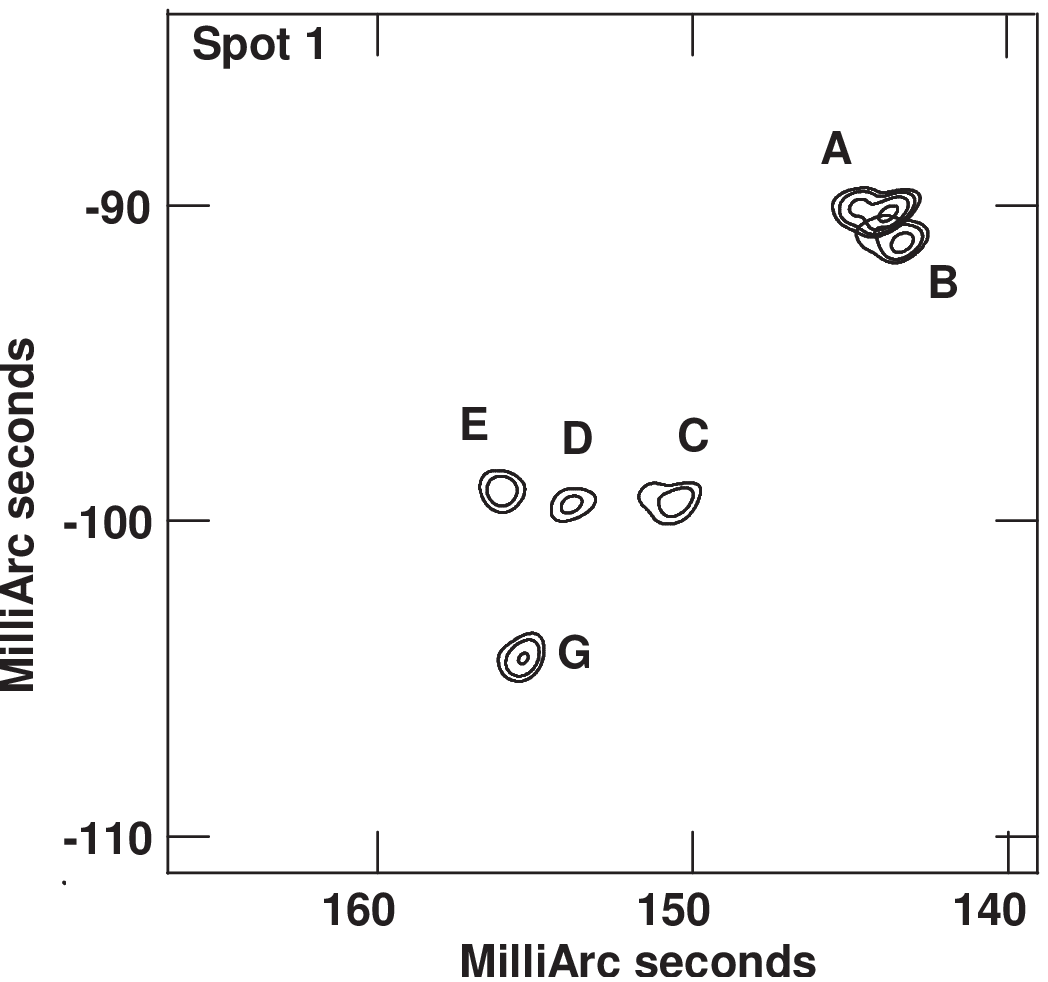}
  \includegraphics[width=5cm]{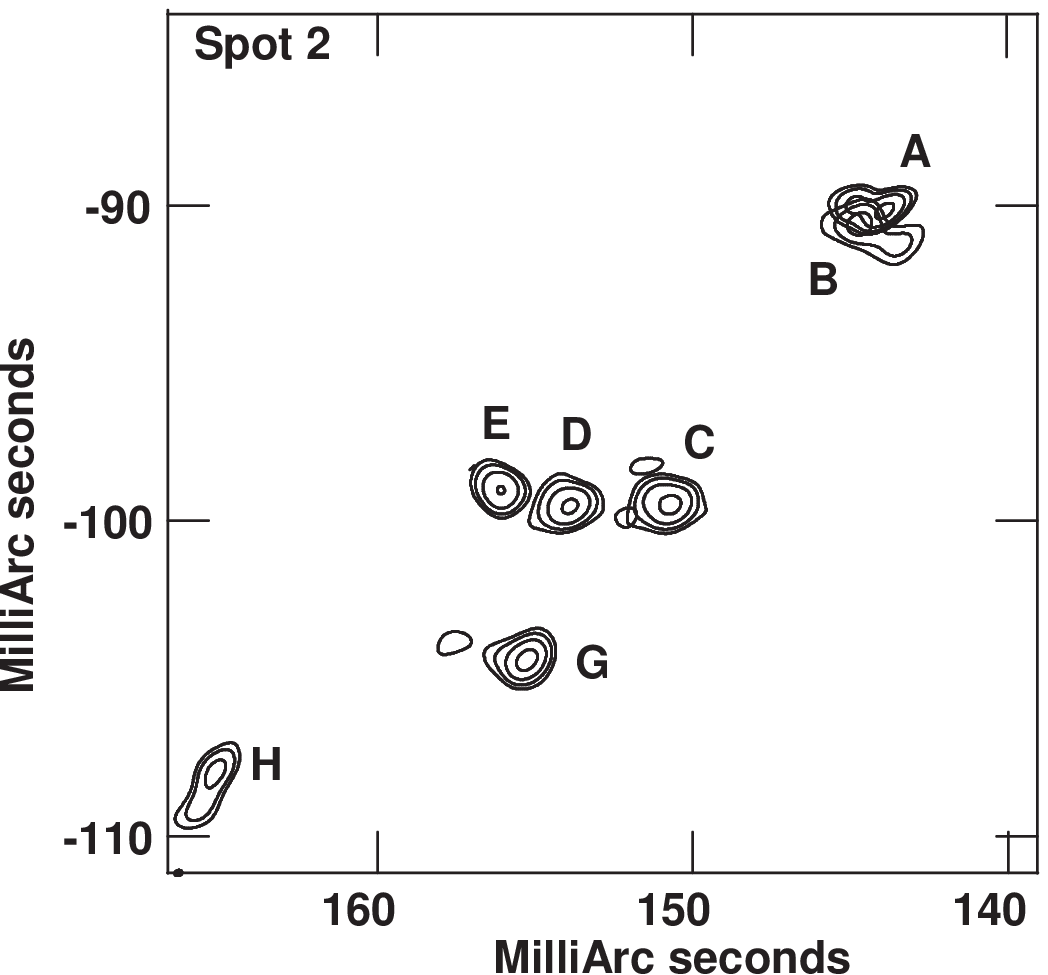}
  \includegraphics[width=5cm]{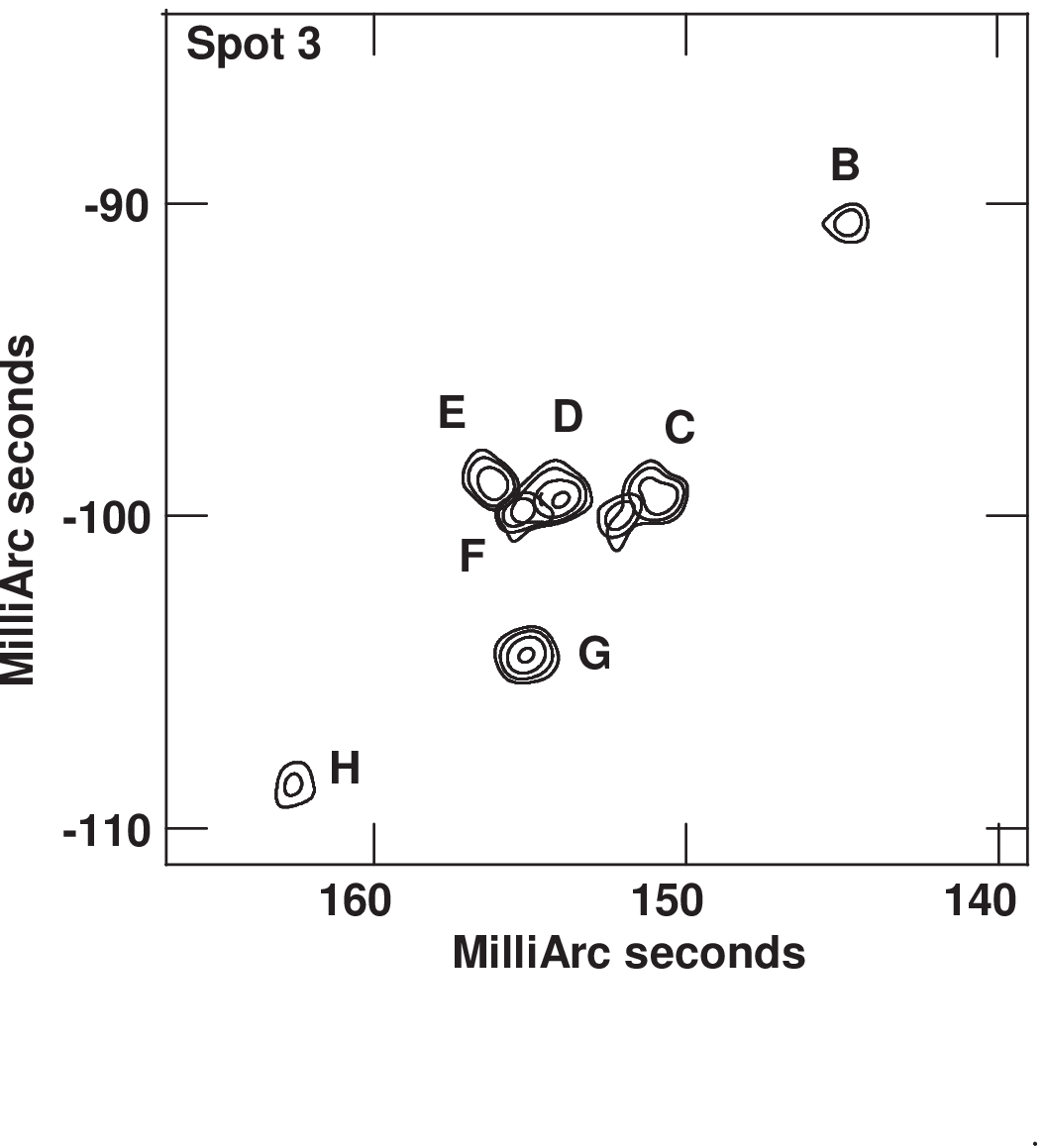}
  \includegraphics[width=5cm]{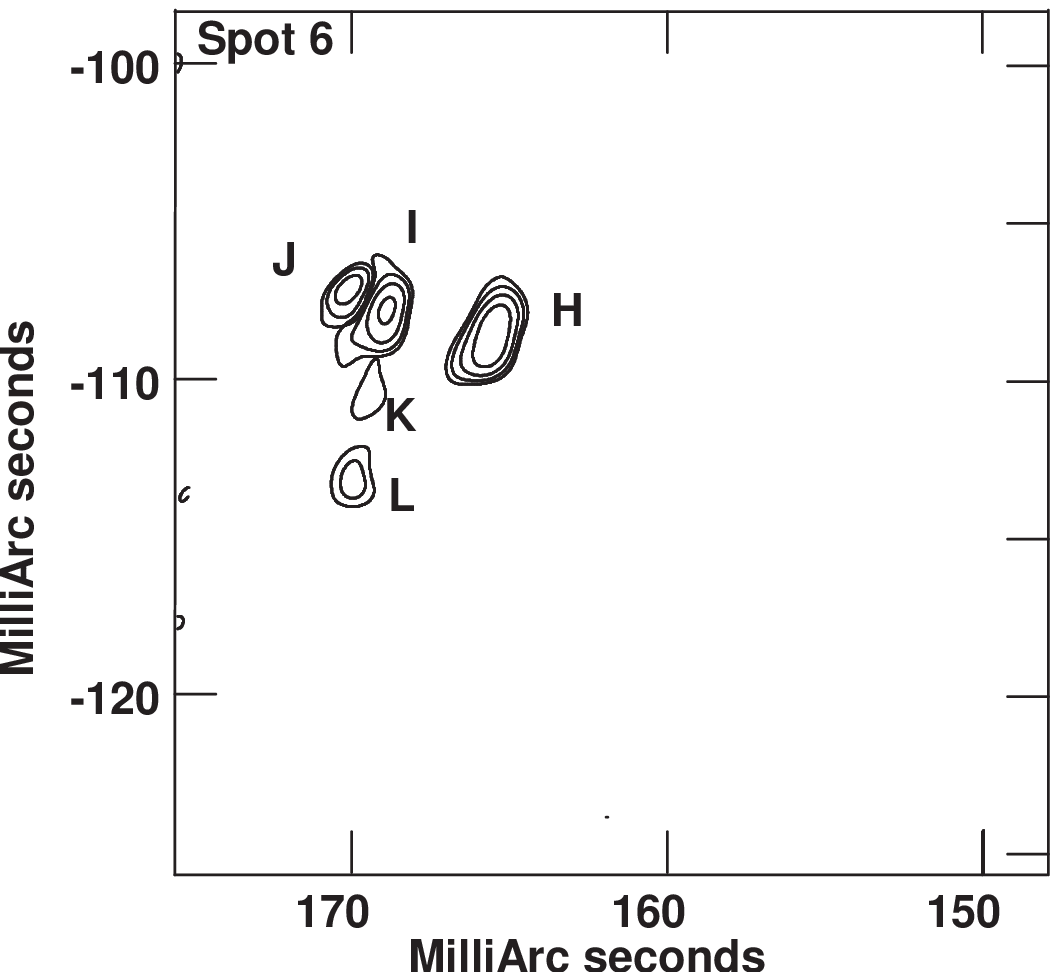}
 \end{center} \vspace{2cm}
 \end{figure}
\clearpage
\begin{figure}
 \begin{center}
  \includegraphics[width=5cm]{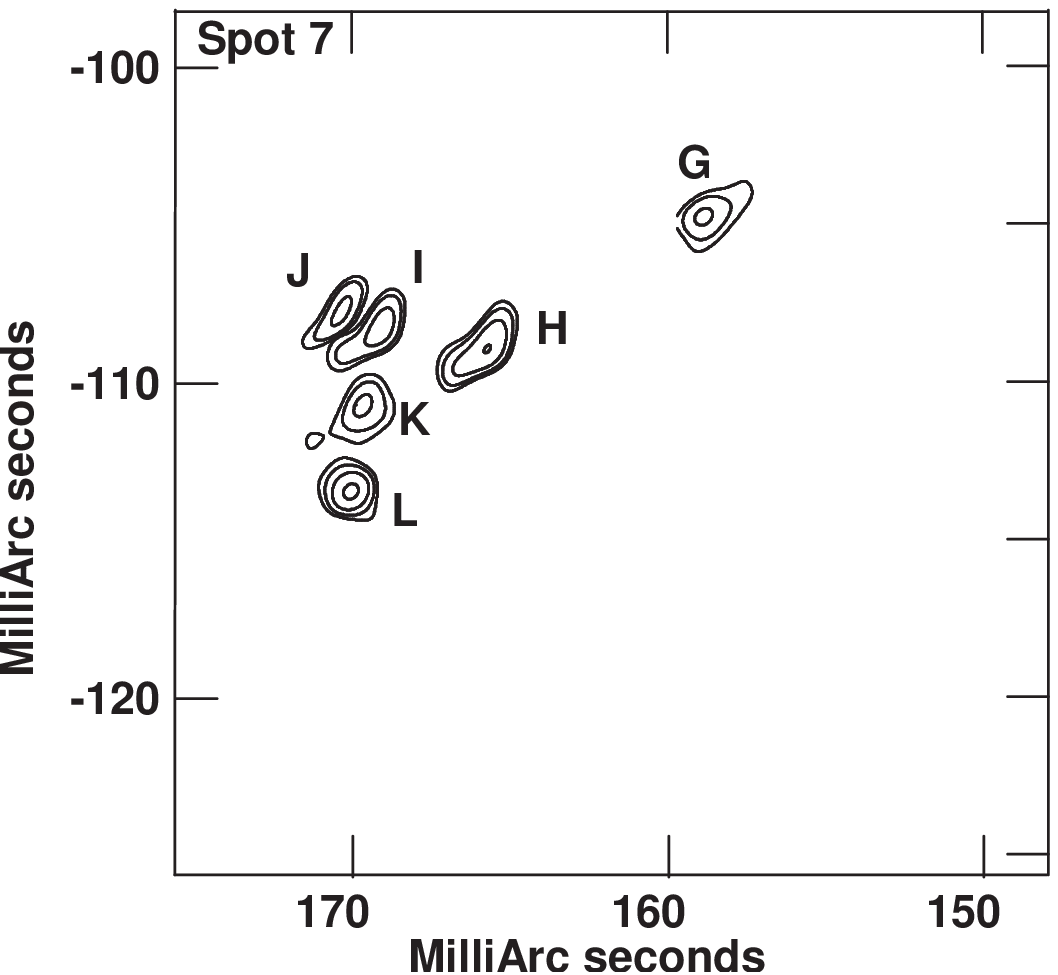}
  \includegraphics[width=5cm]{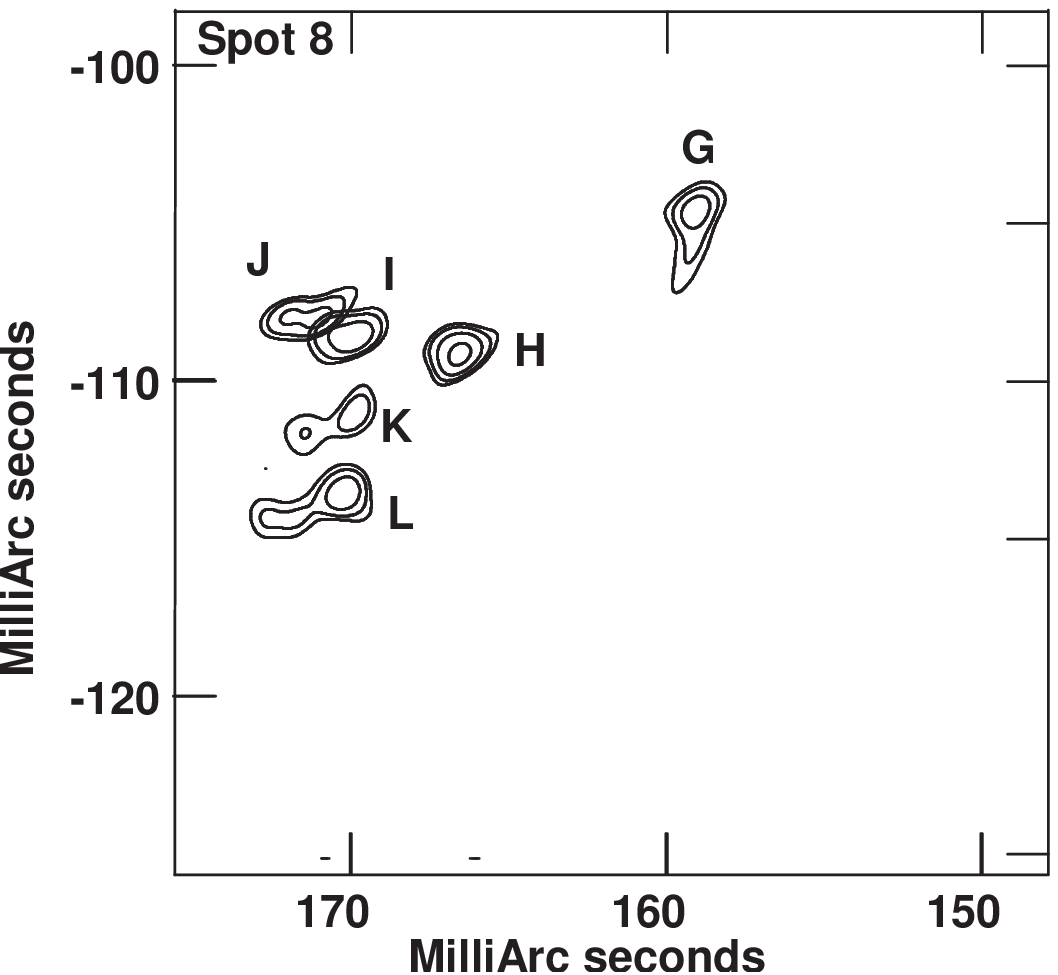}
  \includegraphics[width=5cm]{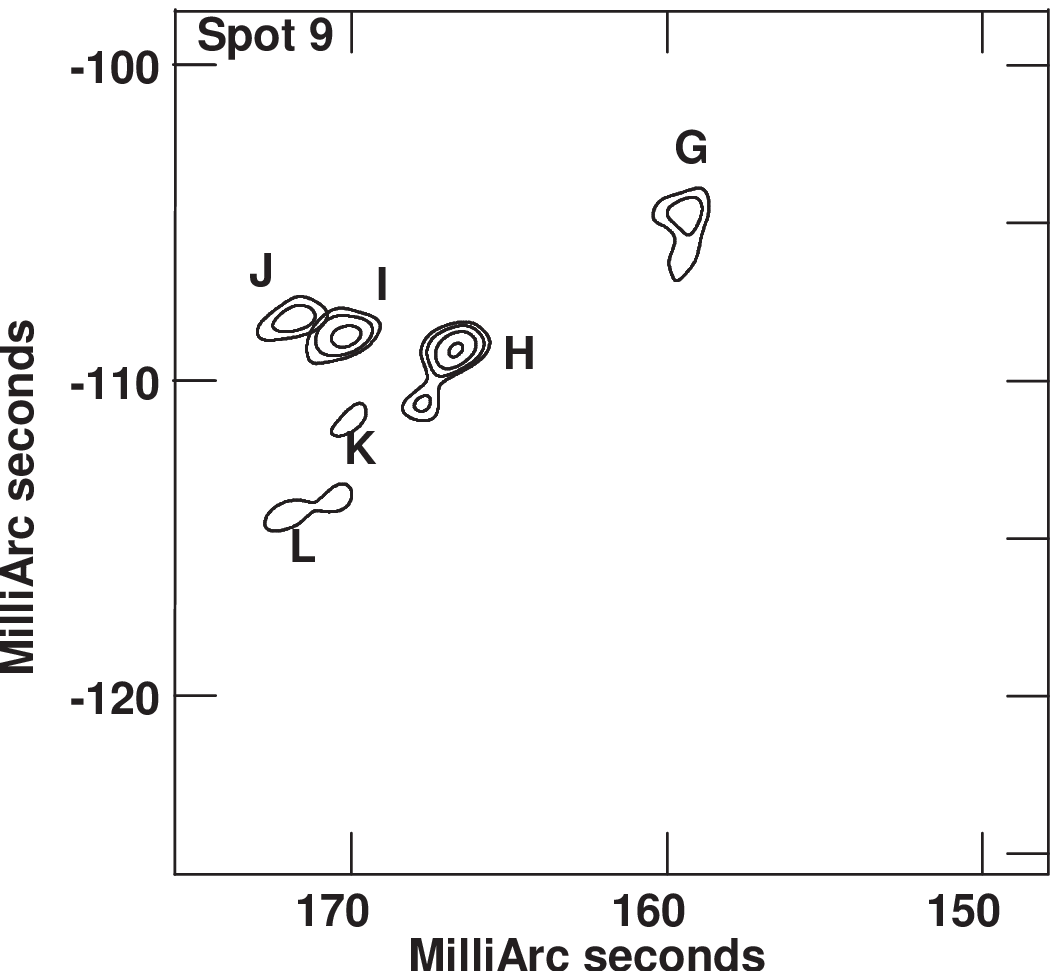}
 \end{center} \vspace{2cm}
 \caption{The position change of  the brightness distribution of 
{\bf Spots} 1, 2, 3, 6, 7, 8, and 9 above 3 $\sigma$. The feature position has been measured with respect to the delay-tracking center in data correlation. 
{\bf The observing epochs are indicated by capital alphabet letters from A to L, see Table 1).}  }\label{fig1}
\end{figure}
\begin{figure}
 \begin{center}
  \includegraphics[width=10cm]{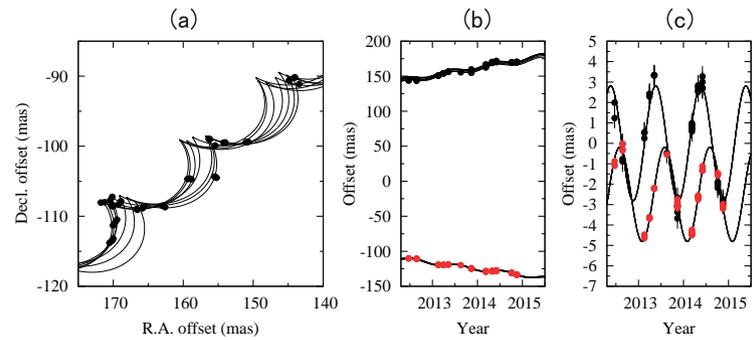}
 \end{center} \vspace{2cm}
 \caption{Motion of seven H$_2$O maser spots 
{\bf used for the annual parallax fitting. }
The solid lines show the best fitting results of parallax and proper motions. (a) Positions in the sky. (b) Time variation of the position for R.A. direction and Dec direction (black and red filled circles, respectively).
(c) Same as (b) except for subtracting the proper motion from the observed motion, allowing the effect of only the parallax to be found. }\label{fig1}
\end{figure}
\begin{figure}
 \begin{center}
  \includegraphics[width=8cm]{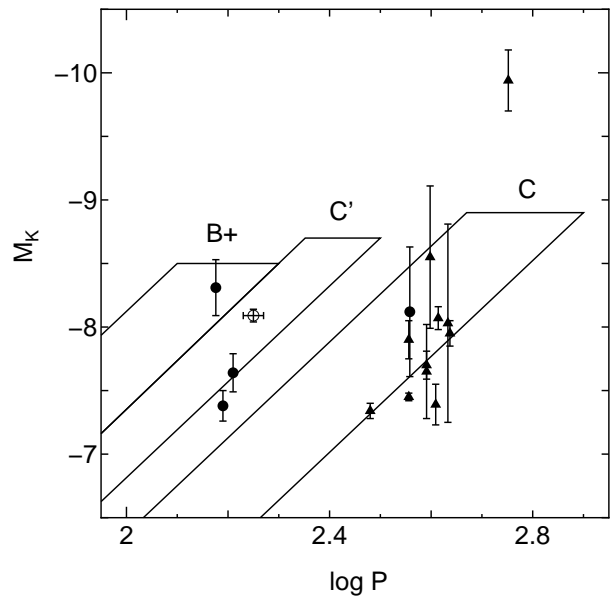}
 \end{center} \vspace{2cm}
 \caption{K'-band magnitude and log $P$ diagram for local Mira variables (filled triangle) and SRVs (filled circle) whose distance was obtained with VLBI astrometry. 
Open circle shows SV Peg. Sequences reproduced from Ita et al. (2004a, b) are also shown. }\label{fig1}
\end{figure}
\begin{figure}
 \begin{center}
  \includegraphics[width=10cm]{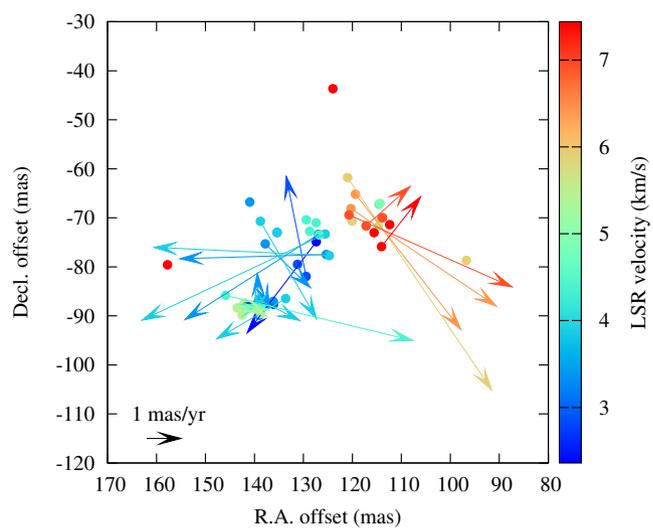}
 \end{center} \vspace{2cm}
 \caption{Distribution and internal motion of the H$_2$O maser spots in SV Peg. }\label{fig1}
\end{figure}

\clearpage



\begin{table}[htbp]
\caption{Observations.} 
\label{tab:J2216+35} 
\begin{center}
\begin{tabular}{rccccc} 
\hline 
Epoch & Date & Synthesized Beam & PA & Number  \\
 & & [mas$\times$mas] & [deg]  & {} \\
(1) & (2) & (3) & (4) & (5) \\
\hline
A & 2012 Jun 20 & $1.30 \times 0.68$&134 & 8 \\
B & 2012 Aug 23 & $1.28 \times 0.81$&144 & 11 \\
C & 2013 Feb 17 & $1.29 \times 0.76$&134 & 12 \\
D & 2013 Mar 30 & $1.35 \times 0.81$&140 & 12\\
E & 2013 May 08 & $1.27 \times 0.95$&162 & 16 \\
F & 2013 Aug 18 & $1.31 \times 0.79$&141 & 10\\
G & 2013 Nov 12 & $1.34 \times 0.88$&144 &  11\\
H & 2014 Mar 09 & $1.28 \times 0.81$&141 &  9\\
I & 2014 Apr 23 & $1.29 \times 0.83$&142 &  5\\
J & 2014 Jun 03 & $1.36 \times 0.68$&132& 5\\
K & 2014 Oct 05 & $1.38 \times 0.85$&144 &  7\\
L & 2014 Nov 16 & $1.25 \times 0.93$&145 & 10\\
M & 2015 Apr 15 & $1.31 \times 0.82$&157 &  4\\
\hline
\end{tabular} 
\begin{tabnote}
(1) Observing epoch, (2) Observing date, (3) Major axis and minor axis of the CLEAN beam, (4) Position angle of the CLEAN beam, (5) Number of detected maser spots.
\end{tabnote}
\end{center} 
\end{table}


\begin{table}[htbp]
\caption{Time variation of the intensity of J2216+3518.} 
\label{tab:J2216+35} 
\begin{center}
\begin{tabular}{rccc} 
\hline 
Epoch & Date &  Peak Intensity & Noise  \\
 &  & [mJy beam$^{-1}$] & [mJy beam$^{-1}$]  \\
(1) & (2) & (3) & (4) \\
\hline
A &  2012 Jun 20 &  331 & 2.7  \\
B & 2012 Aug 23 & 297 & 2.1  \\
C & 2013 Feb 17 & 394 & 1.0  \\
D & 2013 Mar 30 &  397 & 1.2  \\
E & 2013 May 08 &  487 & 1.1  \\
F & 2013 Aug 18 &  417 & 2.5  \\
G & 2013 Nov 12 &  324 & 1.2  \\
H & 2014 Mar 09 &  434 & 1.3  \\
I & 2014 Apr 23 &  420 & 1.2 \\
J & 2014 Jun 03  & 403 & 2.9  \\
K & 2014 Oct 05  & 320 & 1.8  \\
L & 2014 Nov 16  & 351 & 1.4  \\
M & 2015 Apr 15  & 288 & 1.9  \\
\hline
\end{tabular} 
\begin{tabnote}
(1) Observing epoch, (2) Observing date, (3) Peak intensity, (4) Noise level.
\end{tabnote}
\end{center} 
\end{table}


\begin{table}[htbp]
\caption{The fitted parameters of the infrared light curve.} 
\label{tab:fourier-matome} 
\begin{center}
\begin{tabular}{cccccc} 
\hline 
Band &$P$ & $m_0$ & $\Delta m$ & $\theta$ & RMS \\ 
        & [day]  & [mag] & [mag] & [day] &  \\ 
(1) & (2) & (3) & (4) & (5) & (6) \\
\hline 
\multicolumn{6}{l}{Kagoshima 1-m telescope} \\
K' & $177\pm8$ & $-0.48\pm0.01$ & $0.08\pm0.02$ & $46\pm6$ & 0.078 \\
  & $260\pm18$ & $-0.48\pm0.01$ & $0.09\pm0.01$ & $21\pm9$ & 0.073 \\
  & $341\pm25$ & $-0.44\pm0.02$ & $0.12\pm0.02$ & $224\pm8$ & 0.066 \\
H & $177^*$ & $-0.07\pm0.04$ & $0.24\pm0.08$ & $42\pm5$ &--- \\
J & $177^*$ & $0.90\pm0.04$ & $0.21\pm0.07$ & $41\pm6$ & --- \\
\hline 
\multicolumn{6}{l}{Hipparcos} \\
V & $172\pm13$ & $7.89\pm0.01$ & $0.24\pm0.01$ & $118\pm1$ & 0.068 \\
\hline
\end{tabular}
\begin{tabnote}
(1) Observing band, (2) period, (3) averaged magnitude, (4) amplitude, (5) phase, (6) minimum local RMS value. * Fixed value. 
\end{tabnote}
\end{center}
\end{table}

\begin{table}[htbp]
\caption{$V_{\rm LSR}$, positions, intensity, and internal motion of the maser spots in SV Peg.} 
\label{tab:internal-vera_1} 
\begin{center}
\begin{tabular}{rccccccc} 
\hline 
Spot & $V_\mathrm{LSR}$  & R.A. origin & Dec. origin & $S$ & SNR & $\mu_\mathrm{R.A.}^\mathrm{int}$ & $\mu_\mathrm{Dec.}^\mathrm{int}$ \\ 
 & [km s$^{-1}$] & [mas] & [mas] & [Jy Beam$^{-1}$] &  & [mas yr$^{-1}$] & [mas yr$^{-1}$] \\ 
\hline 
1 & 3.21 & $136.11\pm0.88$ & $-87.65\pm0.30$ & 5.97 & 31.9 & $0.84\pm0.91$ & $0.04\pm0.50$ \\
2 & 3.63 & $137.30\pm0.73$ & $-87.42\pm0.24$ & 4.18 & 17.7 & $-0.04\pm0.75$ & $-0.12\pm0.47$ \\
3 & 4.05 & $138.95\pm0.75$ & $-86.79\pm0.29$ & 1.58 & 10.2 & $-1.17\pm0.73$ & $-0.59\pm0.48$ \\
4 & 4.47 & $145.89\pm2.87$ & $-85.83\pm0.14$ & 0.71 & 10.1 & $-5.47\pm1.87$ & $-1.31\pm0.45$ \\
5 & 3.63 & $139.02\pm1.56$ & $-89.88\pm0.98$ & 0.99 & 16.5 & $0.05\pm0.86$ & $1.23\pm0.61$ \\
6 & 4.05 & $139.10\pm0.16$ & $-86.71\pm0.77$ & 5.28 & 63.8 & $0.11\pm0.54$ & $-0.39\pm0.53$ \\
7 & 4.47 & $140.47\pm0.29$ & $-88.28\pm0.42$ & 1.91 & 14.1 & $-0.33\pm0.55$ & $0.09\pm0.47$ \\
8 & 4.89 & $141.79\pm0.74$ & $-87.95\pm0.20$ & 2.31 & 19.6 & $-0.67\pm0.62$ & $-0.13\pm0.45$ \\
9 & 5.31 & $142.06\pm0.96$ & $-87.39\pm0.39$ & 1.00 & 10.1 & $-0.67\pm0.67$ & $-0.39\pm0.47$ \\
10 & 2.79 & $127.37\pm$\ --- & $-74.89\pm$\ --- & 0.60 & 8.3 & $2.01\pm$\ --- & $-2.65\pm$\ --- \\
11 & 3.21 & $129.44\pm0.95$ & $-81.96\pm0.84$ & 1.04 & 13.0 & $0.58\pm0.86$ & $2.92\pm0.74$ \\
12 & 3.63 & $125.34\pm1.56$ & $-77.47\pm2.51$ & 2.40 & 12.3 & $4.27\pm1.49$ & $-0.12\pm2.27$ \\
13 & 4.05 & $124.73\pm0.65$ & $-77.69\pm3.34$ & 2.17 & 15.6 & $5.11\pm0.81$ & $0.25\pm3.11$ \\
14 & 5.73 & $117.23\pm0.85$ & $-71.75\pm0.66$ & 1.18 & 16.5 & --- & --- \\
15 & 6.15 & $120.30\pm$\ --- & $-68.10\pm$\ --- & 1.51 & 18.6 & $-4.25\pm$\ --- & $-2.84\pm$\ --- \\
16 & 6.57 & $120.70\pm3.22$ & $-69.42\pm2.59$ & 0.54 & 7.3 & $-4.76\pm2.05$ & $-2.09\pm1.65$ \\
17 & 7.01 & $114.07\pm$\ --- & $-75.86\pm$\ --- & 3.40 & 20.4 & $-1.15\pm$\ --- & $1.45\pm$\ --- \\
18 & 7.43 & $112.39\pm0.99$ & $-71.38\pm0.78$ & 1.10 & 7.6 & --- & --- \\
19 & 5.73 & $120.97\pm1.43$ & $-61.77\pm3.26$ & 0.56 & 8.4 & $-4.21\pm1.14$ & $-6.19\pm2.34$ \\
20 & 6.15 & $119.36\pm0.96$ & $-65.18\pm2.39$ & 1.37 & 17.7 & $-3.07\pm0.89$ & $-3.93\pm1.82$ \\
21 & 6.57 & $117.17\pm0.62$ & $-71.60\pm1.18$ & 1.71 & 23.7 & $-1.28\pm0.73$ & $1.15\pm1.04$ \\
22 & 7.01 & $115.57\pm0.76$ & $-73.02\pm0.58$ & 0.74 & 10.4 & --- & --- \\
23 & 3.21 & $131.20\pm1.62$ & $-79.48\pm1.30$ & 0.43 & 6.1 & --- & --- \\
24 & 3.63 & $137.80\pm0.74$ & $-75.31\pm1.97$ & 0.64 & 6.9 & $-1.33\pm0.59$ & $-1.27\pm0.79$ \\
25 & 4.05 & $138.80\pm$\ --- & $-70.70\pm$\ --- & 0.46 & 7.6 & $-1.64\pm$\ --- & $-2.84\pm$\ --- \\
26 & 3.63 & $126.93\pm0.81$ & $-73.37\pm0.32$ & 1.23 & 5.4 & $3.89\pm1.15$ & $-2.48\pm0.59$ \\
27 & 4.05 & $125.56\pm$\ --- & $-73.29\pm$\ --- & 1.39 & 8.3 & $5.34\pm$\ --- & $-2.51\pm$\ --- \\
28 & 4.47 & $128.72\pm0.57$ & $-72.74\pm0.41$ & 0.72 & 7.7 & --- & --- \\
29 & 4.05 & $133.61\pm$\ --- & $-86.46\pm$\ --- & 0.99 & 7.9 & $2.01\pm$\ --- & $-1.17\pm$\ --- \\
30 & 4.47 & $126.76\pm0.57$ & $-73.48\pm0.41$ & 0.84 & 9.0 & --- & --- \\

\end{tabular}
\end{center}
\end{table}

\clearpage

\begin{table}[htbp]

\begin{center}
\begin{tabular}{rccccccc} 

31 & 4.47 & $127.42\pm0.57$ & $-71.00\pm0.41$ & 0.60 & 6.5 & --- & --- \\
32 & 4.47 & $129.43\pm0.57$ & $-70.40\pm0.41$ & 0.57 & 6.1 & --- & --- \\
33 & 4.47 & $138.60\pm0.85$ & $-88.10\pm0.66$ & 1.02 & 9.9 & --- & --- \\
34 & 4.47 & $142.27\pm0.85$ & $-87.90\pm0.66$ & 0.99 & 9.6 & --- & --- \\
35 & 5.31 & $143.51\pm0.85$ & $-88.34\pm0.66$ & 0.62 & 8.9 & --- & --- \\
36 & 3.63 & $135.38\pm1.83$ & $-72.98\pm1.48$ & 0.65 & 11.9 & --- & --- \\
37 & 4.05 & $135.35\pm1.83$ & $-72.97\pm1.48$ & 0.38 & 7.4 & --- & --- \\
38 & 4.47 & $114.55\pm0.51$ & $-67.23\pm0.36$ & 1.49 & 8.0 & --- & --- \\
39 & 4.89 & $114.34\pm0.51$ & $-67.05\pm0.36$ & 1.22 & 7.3 & --- & --- \\
40 & 5.31 & $114.61\pm0.52$ & $-71.58\pm0.36$ & 1.41 & 7.4 & --- & --- \\
41 & 5.31 & $49.20\pm0.51$ & $-209.34\pm0.36$ & 1.21 & 8.2 & --- & --- \\
42 & 5.73 & $120.07\pm0.51$ & $-70.66\pm0.36$ & 1.53 & 9.2 & --- & --- \\
43 & 7.01 & $123.99\pm0.99$ & $-43.66\pm0.77$ & 1.46 & 7.2 & --- & --- \\
44 & 6.15 & $114.03\pm1.26$ & $-70.02\pm1.01$ & 0.39 & 8.1 & --- & --- \\
45 & 6.57 & $113.86\pm1.26$ & $-69.96\pm1.01$ & 0.42 & 8.3 & --- & --- \\
46 & 4.89 & $138.79\pm1.63$ & $-89.49\pm1.30$ & 0.56 & 8.1 & --- & --- \\
47 & 5.31 & $138.67\pm1.62$ & $-89.49\pm1.30$ & 0.34 & 5.8 & --- & --- \\
48 & 3.63 & $136.17\pm1.62$ & $-87.05\pm1.30$ & 0.47 & 7.0 & --- & --- \\
49 & 7.01 & $157.75\pm1.33$ & $-79.58\pm0.95$ & 1.09 & 6.0 & --- & --- \\
50 & 2.37 & $177.18\pm1.33$ & $-33.25\pm0.94$ & 0.73 & 5.2 & --- & --- \\
51 & 5.31 & $142.49\pm1.20$ & $-89.83\pm0.82$ & 0.37 & 5.5 & --- & --- \\
52 & 3.63 & $140.96\pm1.79$ & $-66.79\pm1.37$ & 0.56 & 5.6 & --- & --- \\
53 & 5.73 & $96.72\pm1.79$ & $-78.66\pm1.37$ & 0.51 & 5.2 & --- & --- \\
\hline 
\end{tabular}
\end{center}
\end{table}

\clearpage


\begin{table}[htbp]
\caption{Position offset of the seven maser spots available for the measurement of the annual parallax.} 
\label{tab:ap} 
\begin{center}
\begin{tabular}{rcc} 
\hline 
Epoch & R.A. Offset  & Dec. Offset      \\
         & [mas]         & [mas]              \\ 
\hline 
\multicolumn{3}{l}{Spot 1} \\
A	&	144.03 	$\pm$	0.51 	&	$-$90.16 	$\pm$	0.16 	\\
B	&	143.43 	$\pm$	0.51 	&	$-$91.13 	$\pm$	0.16 	\\
C	&	150.71 	$\pm$	0.51 	&	$-$99.38 	$\pm$	0.16 	\\
D	&	153.98 	$\pm$	0.51 	&	$-$99.47 	$\pm$	0.16 	\\
E	&	156.19 	$\pm$	0.51 	&	$-$99.00 	$\pm$	0.16 	\\
G	&	155.53 	$\pm$	0.51 	&	$-$104.34 	$\pm$	0.16 	\\
\hline 
\multicolumn{3}{l}{Spot 2} \\						
A	&	144.07 	$\pm$	0.51 	&	$-$90.21 	$\pm$	0.16 	\\
B	&	144.81 	$\pm$	0.51 	&	$-$90.66 	$\pm$	0.16 	\\
C	&	150.89 	$\pm$	0.50 	&	$-$99.44 	$\pm$	0.16 	\\
D	&	154.06 	$\pm$	0.50 	&	$-$99.50 	$\pm$	0.16 	\\
E	&	156.22 	$\pm$	0.50 	&	$-$98.99 	$\pm$	0.16 	\\
F	&	155.40 	$\pm$	0.50 	&	$-$104.38 	$\pm$	0.16 	\\
G	&	163.00 	$\pm$	0.50 	&	$-$108.49 	$\pm$	0.17 	\\
\hline 
\multicolumn{3}{l}{Spot 3} \\						
B	&	144.89 	$\pm$	0.51 	&	$-$90.61 	$\pm$	0.16 	\\
C	&	151.01 	$\pm$	0.50 	&	$-$99.41 	$\pm$	0.16 	\\
D	&	154.21 	$\pm$	0.50 	&	$-$99.49 	$\pm$	0.16 	\\
E	&	156.34 	$\pm$	0.50 	&	$-$98.96 	$\pm$	0.16 	\\
F	&	155.42 	$\pm$	0.51 	&	$-$99.90 	$\pm$	0.16 	\\
G	&	155.23 	$\pm$	0.50 	&	$-$104.51 	$\pm$	0.16 	\\
H	&	162.61 	$\pm$	0.51 	&	$-$108.70 	$\pm$	0.17 	\\
\hline 
\multicolumn{3}{l}{Spot 6} \\					
H	&	165.53 	$\pm$	0.50 	&	$-$108.59 	$\pm$	0.16 	\\
I	&	168.91 	$\pm$	0.50 	&	$-$107.85 	$\pm$	0.16 	\\
J	&	170.15 	$\pm$	0.50 	&	$-$107.23 	$\pm$	0.16 	\\
K	&	169.47 	$\pm$	0.51 	&	$-$110.47 	$\pm$	0.20 	\\
L	&	169.99 	$\pm$	0.51 	&	$-$113.20 	$\pm$	0.18 	\\

\end{tabular}
\end{center}
\end{table}

\clearpage

\begin{table}[htbp]

\begin{center}
\begin{tabular}{rcc}

\hline 
\multicolumn{3}{l}{Spot 7} \\				
G	&	158.83 	$\pm$	0.51 	&	$-$104.72 	$\pm$	0.16 	\\
H	&	165.81 	$\pm$	0.50 	&	$-$108.86 	$\pm$	0.16 	\\
I	&	169.19 	$\pm$	0.50 	&	$-$108.26 	$\pm$	0.16 	\\
J	&	170.43 	$\pm$	0.51 	&	$-$107.73 	$\pm$	0.16 	\\
K	&	169.68 	$\pm$	0.51 	&	$-$110.80 	$\pm$	0.16 	\\
L	&	170.12 	$\pm$	0.50 	&	$-$113.40 	$\pm$	0.16 	\\
\hline 
\multicolumn{3}{l}{Spot 8} \\									
G	&	159.09 	$\pm$	0.50		&	$-$104.61 	$\pm$	0.16 	\\
H	&	166.55 	$\pm$	0.50 	&	$-$109.07 	$\pm$	0.16 	\\
I	&	170.02 	$\pm$	0.51 	&	$-$108.58 	$\pm$	0.16 	\\
J	&	171.23 	$\pm$	0.51 	&	$-$108.00 	$\pm$	0.16 	\\
K	&	169.88 	$\pm$	0.51 	&	$-$111.06 	$\pm$	0.16 	\\
L	&	170.22 	$\pm$	0.50		&	$-$113.54 	$\pm$	0.17 	\\
\hline 
\multicolumn{3}{l}{Spot 9} \\									
G	&	159.40 	$\pm$	0.51 	&	$-$104.68 	$\pm$	0.16 	\\
H	&	166.67 	$\pm$	0.51 	&	$-$108.99 	$\pm$	0.16 	\\
I	&	170.21 	$\pm$	0.51 	&	$-$108.53 	$\pm$	0.16 	\\
J	&	171.78 	$\pm$	0.51 	&	$-$108.06 	$\pm$	0.16 	\\
K	&	170.04 	$\pm$	0.51 	&	$-$111.31 	$\pm$	0.19 	\\
L	&	170.53 	$\pm$	0.51 	&	$-$113.78 	$\pm$	0.18 	\\

\hline
\end{tabular}
\begin{tabnote}

\end{tabnote}
\end{center}
\end{table}

\end{document}